\documentclass[aps,pra,preprint,groupedaddress,showpacs]{revtex4-1}
\usepackage{graphicx}
\begin{document}

\title{Stability of flip and exchange symmetric entangled state classes
under invertible local operations}

\author{Z. Gedik}

\email[]{gedik@sabanciuniv.edu}

\affiliation{Faculty of Engineering and Natural Sciences, Sabanci
University, Tuzla, Istanbul 34956, Turkey}

\date{\today}

\begin{abstract}
Flip and exchange symmetric (FES) many-qubit states, which can be
obtained from a state with the same symmetries by means of
invertible local operations (ILO), constitute a set of curves in the
Hilbert space. Eigenstates of FES ILOs correspond to vectors that
cannot be transformed to other FES states. This means equivalence
classes of states under ILO can be determined in a systematic way
for an arbitrary number of qubits. More important, for entangled
states, at the boundaries of neighboring equivalence classes, one
can show that when the fidelity between the final state after an ILO
and a state of the neighboring class approaches unity, probability
of success decreases to zero. Therefore, the classes are stable
under ILOs.
\end{abstract}


\maketitle

For composite systems, quantum states which cannot be written as a
product of the states of individual subsystems have been recognized
since the early days of quantum mechanics \cite{1,2,3}. This
phenomenon, called entanglement, is at the center of quantum
information theory since it is a key resource in quantum processes
\cite{4}. A natural class of operations suitable for manipulating
entanglement is that of local operations and classical communication
(LOCC) \cite{5,6}. For probabilistic transformations, the condition
of certainty can be removed to allow conversion of the states
through stochastic local operations and classical communication
(SLOCC) \cite{7}. This coarse graining simplifies the equivalence
classes labeled by continuous parameters in case of the local
unitary operations. Two states are equivalent under SLOCC if an
invertible local operation (ILO) relating them exists \cite{8}. For
example, in case of three qubits, $\vert\psi\rangle$ and
$\vert\phi\rangle$ are in the same equivalence class if
$\vert\phi\rangle=A\otimes B\otimes C\vert\psi\rangle$ where $A$,
$B$ and $C$ are invertible operators corresponding to each party.

All entangled pure states of two qubits can be converted to the
Einstein-Podolsky-Rosen (EPR)
$(1/\sqrt{2})(\vert00\rangle+\vert11\rangle)$ state under SLOCC
\cite{8,9}. In other words, there is a single equivalence class. For
three qubits, by calculating transformed states explicitly, it has
been shown that there are two inequivalent states ($\vert
GHZ\rangle$ and $\vert W\rangle$) \cite{8}. Starting from four
qubits, equivalence classes of multipartite systems are labeled by
at least one continuous parameter \cite{8}. Using the isomorphism
$SU(2)\otimes SU(2)\simeq SO(4)$ from Lie group theory, nine
different ways of entanglement have been found for four qubits
\cite{10}. There are no complete classifications for five or more
qubits. However, for exchange symmetric $n-$qubit states,
entanglement classification under SLOCC has been achieved by
introducing two parameters called diversity degree and degeneracy
configuration. Also, the number of families has been shown to grow
as the partition function of the number of qubits \cite{11}.

Flip and exchange symmetric (FES) many-qubit states are those having
both permutation (exchange) and 0-1 (flip) symmetry. Namely, FES
sates are invariant when two qubits are interchanged or when all 0s
(1s) are changed to 1s (0s). In the last few years, there has been
extensive studies on the entanglement properties of symmetric
multipartite states \cite{12,13}. The main reason for utilization of
FES states is the simplicity of the form of their entanglement
classes under SLOCC. Since any linear combination of two FES states
is also FES, they form a subspace whose dimension is $\lfloor
n/2\rfloor+1$ where $\lfloor.\rfloor$ denotes integer part function.
Therefore, for large $n$ values dimension is approximately $n/2$.
The main purpose of the present work is to study stability of SLOCC
equivalence classes under ILO in the FES subspace. Such a
restriction may seem to be an oversimplification when compared to
the $2^n$ dimensional Hilbert space problem. However, FES states are
important if one considers for example bosonic qubits where exchange
symmetry is essential. The classification method to be presented is
distinguished in two ways: It is systematic and it gives information
about neighboring equivalence classes and their relative sizes. More
important, it justifies the SLOCC classification in the sense that
at the boundaries of different classes, probabilities to end up with
states arbitrarily close to the states of neighboring classes are
shown to be vanishing.

\textbf{Definition:} $\vert\psi\rangle$ is a FES $n-$qubit state if
it satisfies $X^{\otimes n}\vert\psi\rangle=\vert\psi\rangle$ where
\begin{eqnarray}
X = \left(\begin{array}{cc}
0 & 1 \\
1 & 0 \\
\end{array}\right)
\end{eqnarray}
is the qubit flip operator and
$P_{ij}\vert\psi\rangle=\vert\psi\rangle$. Here $P_{ij}$ is the
exchange operator for the $i^{th}$ and the $j^{th}$ qubits.

The three qubit state $(\vert
GHZ\rangle=(1/\sqrt{2})(\vert000\rangle+\vert111\rangle)$ is FES,
and $(\vert
W\rangle=(1/\sqrt{3})(\vert100\rangle+\vert010\rangle+\vert001\rangle)$
can be made FES with a simple ILO. How near $\vert GHZ\rangle-$type
and $\vert W\rangle-$type states can be is one of the questions
addressed in this study.

Imposing exchange symmetry greatly simplifies the classification
problem. All local operators become identical and thus will be
denoted by $M$ \cite{14}. Qubit flip symmetry reduces the number of
parameters, i.e., entries of the $2\times2$ matrix $M$, from four to
two. Hence, FES ILOs can be written as
\begin{eqnarray}
M = \left(\begin{array}{cc}
a & b \\
b & a \\
\end{array}\right),
\end{eqnarray}
where $a^2\neq b^2$. Since $a$ and $b$ cannot simultaneously vanish,
$M$ can be simplified further through division by $a$ or $b$. For
$a\neq 0$
\begin{eqnarray}
M (t)= f(t)\left (\begin{array}{cc}
1 & t \\
t & 1 \\
\end{array}\right),
\end{eqnarray}
where $t^2\neq 1$, and the function $f(t)\neq0$ will be shown to be
bounded. All of the results of the current work remain the same for
the $b\neq 0$ case where the diagonal and the anti-diagonal of
$M(t)$ are simply interchanged.

Let $\vert\psi(0)\rangle$ be a normalized arbitrary $n$ qubit  FES
state. All equivalent normalized states can then be written as
\begin{equation}
\vert\psi(t)\rangle=\frac{M^{\otimes n}(t)\vert\psi(0)\rangle}
{\sqrt{\langle\psi(0)\vert(M^\dagger M)^{\otimes
n}\vert\psi(0)\rangle}},
\end{equation}
which are FES as well. They lie on a curve parametrized by $t$
provided that $t$ is real. As $t$ changes from $-\infty$ to
$\infty$, excluding $\pm1$, $\vert\psi(t)\rangle$ traces the curve.
However, if $\vert\psi(0)\rangle$ turns out to be an eigenstate of
$M^{\otimes n}(t)$, no ILO will alter it or by definition
$\vert\psi(0)\rangle$ will form an equivalence class by itself.
Eigenstates of $M^{\otimes n}$ are of the form
$\bigotimes_{k=1}^{n}\vert\pm\rangle_k$ where
$\vert\pm\rangle=(1/\sqrt{2})(\vert0\rangle\pm\vert1\rangle)$, and
number of $\vert+\rangle$ and $\vert-\rangle$ states in the
Kronecker product are $p$ and $q=n-p$, respectively. Flip symmetric
ones are those with even $q$. Eigenvalues are given by
\begin{equation}
\lambda_{pq}=f^n(t)(1+t)^p (1-t)^q , \label{lambda}
\end{equation}
and they are $n!/p!q!$  fold degenerate.

\textbf{Definition:} The eigenstate $\vert\psi_{pq}\rangle$ denotes
the FES state obtained by evaluating the symmetric linear
combination of degenerate eigenstates corresponding to eigenvalue
$\lambda_{pq}$ given by eq. (\ref{lambda}).

For example, four qubit $\vert\psi_{pq}\rangle$ states are given by
\begin{eqnarray}
\begin{array}{rr}
\vert\psi_ {40}\rangle= & \vert ++++\rangle\\
\vert\psi_ {22}\rangle= &\frac{1}{\sqrt{6}} (\vert ++--\rangle+\vert -+-+\rangle+\vert +--+\rangle\\
&\vert --++\rangle+\vert +-+-\rangle+\vert -++-\rangle)\\
\vert\psi_ {04}\rangle= & \vert ----\rangle .
\end{array}
\end{eqnarray}

There is a basis of the symmetric subspace in terms of Dicke states
\cite{15,16}. One can label these states according to the number of
0's as
\begin{equation}
\vert
S(n,k)\rangle\equiv\sqrt{\frac{k!(n-k)!}{n!}}\sum_{permutations}
\vert \underbrace{0...0}_k\underbrace{1...1}_{n-k}\rangle
\end{equation}
and it is easy to see that
\begin{equation}
\vert \psi_{pq}\rangle=H^{(p+q)}\vert S(p+q,p)\rangle
\end{equation}
where $H$ is the Hadamard matrix
\begin{eqnarray}
H= \frac{1}{\sqrt{2}}\left (\begin{array}{cc}
1 & 1 \\
1 & -1 \\
\end{array}\right).
\end{eqnarray}

In the context of geometric measure of entanglement, it has been
shown that the closest product state to any symmetric multi-qubit
state is necessarily symmetric \cite{17}. $\vert+\rangle^{\otimes
n}$ is the only FES product state for odd $n$. For even $n$,
$\vert-\rangle^{\otimes n}$ is also FES. One can show that, among
the symmetric product states of the form
\begin{equation}
(\cos\theta\vert 0\rangle+\sin\theta\vert 1\rangle)^{\otimes n},
\end{equation}
$\theta=0$ and $\theta=\pi/2$, which corresponds to
$\vert0\rangle^{\otimes n}$ and $\vert1\rangle^{\otimes n}$,
respectively, are the closest ones to the generalized GHZ state
given by
\begin{equation}
\frac{\vert 0\rangle^{\otimes n}+\vert 1\rangle^{\otimes n}}
{\sqrt{2}}.
\end{equation}
Therefore, the closest product state to a generalized GHZ state is
not necessarily FES.

In practice, the only constraint on $M$ is $(M^\dagger M)^{\otimes
n} /\langle\psi(0)\vert(M^\dagger M)^{\otimes
n}\vert\psi(0)\rangle\leq I$ if $M$ is to come from a positive
operator-valued measure. Here the denominator gives the probability
of obtaining the final state $\vert\psi(t)\rangle$ from
$\vert\psi(0)\rangle$. For a given set of transformations $\{M_i\}$
leading to different final states, the normalization condition is
$\sum_i M_i^\dagger M_i= I$ with the immediate consequence $\vert
f(t)\vert\leq1$.

For two qubits, possible even $q$ values are 0 and 2. Both are
nondegenerate and hence the corresponding eigenstates are separable.
Therefore, there are no entangled FES states unreachable from the
EPR state by means of ILO, which is a very well known result
\cite{9}.

In case of three qubits, allowed $q$ values are the same as above,
but this time while $\lambda_{30}$ corresponds to a separable state
$\vert S\rangle=\vert+++\rangle$, $\lambda_{12}$ is threefold
degenerate and it is easy to see that the eigenvector
$\vert\psi_{12}\rangle=(1/\sqrt{3})(\vert +--\rangle+\vert
-+-\rangle+\vert --+\rangle)$ is equivalent to the $\vert W\rangle$
state. Hence, $\vert W\rangle$ is distinguished from other
three-qubit entangled states, for example from $\vert GHZ\rangle$,
in that it is unreachable via ILO which was again noticed earlier
\cite{8}. Having real expansion coefficients in the computational
basis, $\vert GHZ\rangle$ can be written as $\vert
GHZ\rangle=\cos\theta\vert\psi_{12}\rangle+\sin\theta\vert\psi_{30}\rangle$
with $\theta=\pi/6$. In other words, $\vert GHZ\rangle$ lies on the
geodesic connecting the separable $\vert S\rangle$ state and the
$\vert W\rangle$ state transformed into FES form by means of ILOs.
For $\vert\psi(0)\rangle=\vert GHZ\rangle$, $\vert\psi(t)\rangle$ is
again on the same geodesic and approaches the FES $\vert W\rangle$
state when $t\rightarrow -1$, where the probability $\langle
GHZ\vert(M^\dagger M)^{\otimes 3}\vert GHZ\rangle=\vert
f(t)\vert^6[(1+t^2)^3+8t^3]$ goes to zero. Such a tradeoff between
fidelity and conversion probability was pointed out and
experimentally implemented by Walther \emph{et al}. \cite{18}. On
the other hand, $\vert\psi(t)\rangle$ tends to $\vert S\rangle$ as
$t\rightarrow 1$. Thus, almost all FES three qubit states are
equivalent to $\vert GHZ\rangle$ under ILO while $\vert W\rangle$
and $\vert S\rangle$ are two neighbors of this equivalence class.
GHZ-equivalent states are stable under ILO in the sense that the
nearer the final state $\vert\psi(t)\rangle$ to $\vert W\rangle$,
the less is the probability of success. For a graphical
representation see Figure~1.

Allowed $q$ values for four qubits are 0, 2 and 4. The first and the
third are separable $\vert\psi_{40}\rangle$ and
$\vert\psi_{04}\rangle$ states, respectively. The only interesting
one is $\vert\psi_{22}\rangle$ which is nothing but $G_{0,-1,0,1}$
in the notation of ref.~\cite{10} where $G_{abcd}$ is defined by
\begin{eqnarray}
\begin{array}{cc}
G_{abcd}= & \frac{a+d}{2}(\vert 0000\rangle+\vert 1111\rangle)
+ \frac{a-d}{2}(\vert 0011\rangle+\vert 1100\rangle)\\
& +\frac{b+c}{2}(\vert 0101\rangle+\vert 1010\rangle) +
\frac{b-c}{2}(\vert 0110\rangle+\vert 1001\rangle).
\end{array}
\end{eqnarray}
Since there are three different eigenstates, the FES subspace is a
sphere. It is possible to show that all curves start and end on
$\vert\psi_{40}\rangle$ and $\vert\psi_{04}\rangle$, and they either
do not pass through $\vert\psi_{22}\rangle$ or the probability
decays to zero as $\vert\psi(t)\rangle$ approaches
$\vert\psi_{22}\rangle$. The four qubit problem is the simplest
non-trivial case in the sense that there is more than one curve;
there are in fact infinitely many curves. Among the nine classes of
four qubit states, the only FES one is $G_{abcd}$ with $b=a-d$ and
$c=0$, and it corresponds to a great circle on the sphere passing
through $\vert\psi_{22}\rangle$ and making equal angles with
$\vert\psi_{40}\rangle$ and $\vert\psi_{04}\rangle$ \cite{10}. Thus,
all FES four qubit states are equivalent to a $G_{a,a-d,0,d}$ state
under ILO and $G_{0,-1,0,1}$ is a distinguished entangled state in
that it is unreachable starting from the ones with $a\neq0$.
$G_{a,a-d,0,d}$ corresponds to the canonical form $(\vert
GHZ_4\rangle+\mu\vert D_4^{(2)}\rangle)/\sqrt{1+\vert\mu\vert^2}$ of
ref.~\cite{11} where$\vert D_4^{(2)}\rangle$ is the four qubit Dicke
state with two $\vert 0 \rangle$ components and
$\mu=\sqrt{3}(a-d)/(a+d)$. For a graphical representation see
Figure~1.

\begin{figure}[b]
\centering{\includegraphics[width=3in]{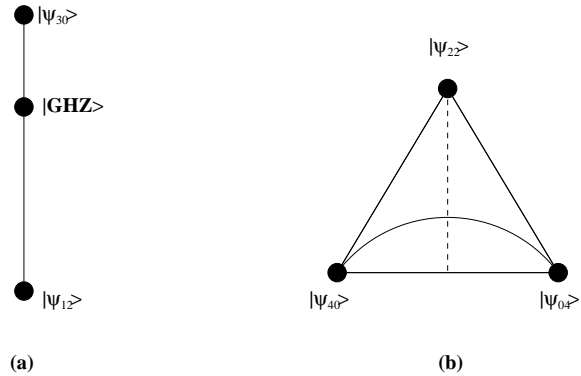}} \caption{Three and
four qubit FES states under ILOs. \textbf{(a)} Almost all FES three
qubit states are equivalent to $\vert GHZ\rangle$ under ILO while
$\vert W\rangle$ and $\vert S\rangle$ are two neighbors of this
equivalence class. \textbf{(b)} All curves start and end on
$\vert\psi_{40}\rangle$ and $\vert\psi_{04}\rangle$, and they either
do not pass through $\vert\psi_{22}\rangle$ or the probability
decays to zero as $\vert\psi(t)\rangle$ approaches
$\vert\psi_{22}\rangle$. The dashed line denotes $G_{a,a-d,0,d}$
states and $\vert\psi_{22}\rangle$ corresponds to $a=0$.}
\end{figure}

In the case of five qubits, there are two entangled FES eigenstates:
$\vert\psi_{32}\rangle$ and $\vert\psi_{14}\rangle$. On the other
hand, $\vert\psi_{42}\rangle$ and $\vert\psi_{24}\rangle$ are two
SLOCC inequivalent six qubit entangled states. The five qubit case
is the same as the four qubit problem in the sense that both are
three dimensional spaces. In general an odd number qubit problem is
equivalent to one less even case. For example, it is possible to see
without any calculation that $\vert\psi_{32}\rangle$ is a FES
entangled state which cannot be reached by ILOs starting from the
others. Furthermore, the curves join $\vert\psi_{50}\rangle$ to
$\vert\psi_{14}\rangle$, but conversion probability decreases to
zero as $\vert\psi(t)\rangle$ tends to $\vert\psi_{14}\rangle$. In
case of six qubits, curves extend from $\vert\psi_{60}\rangle$ to
$\vert\psi_{06}\rangle$. They either do not pass through
$\vert\psi_{42}\rangle$ and $\vert\psi_{24}\rangle$ or if they reach
these two  states, probability for such processes vanish.

For $n$ qubits, the most general FES state can be written as
\begin{equation}
\vert\psi(0)\rangle=\sum_{q=0}^{2\lfloor n/2\rfloor}
c_{pq}\vert\psi_{pq}\rangle
\end{equation}
where the sum runs over even $q$ so that there will be $\lfloor
n/2\rfloor+1$ terms and $\sum\vert c_{pq}\vert^2=1$. Here,$\lfloor
x\rfloor$ stands for the largest integer less than or equal to $x$.
One or two of $\vert\psi_{pq}\rangle$ states will be separable, for
odd and even $n$, respectively. Application of a FES ILO gives
\begin{equation}
\vert\psi(t)\rangle\propto\sum c_{pq}(1+t)^p (1-t)^q
\vert\psi_{pq}\rangle \label{curve}
\end{equation}
which lies on the sphere $S^{\lfloor n/2\rfloor}$. On the surface of
the sphere, a subset described by $\lfloor n/2\rfloor-1$ free
parameters will be enough to obtain all FES states by means of ILOs.
Clearly, provided that $c_{0n}c_{n0}\neq0$,
$\lim_{t\rightarrow1}\vert\psi(t)\rangle=\vert\psi_{n0}\rangle$ and
$\lim_{t\rightarrow-1}\vert\psi(t)\rangle=\vert\psi_{n-2\lfloor
n/2\rfloor,2\lfloor n/2\rfloor}\rangle$, the latter being an
entangled state for odd $n$. A representative subset can be obtained
by taking one point from each such curve connecting the two states.
For example, in case of four qubits this is achieved by
$G_{a,a-d,0,d}$ which cuts all the curves from
$\vert\psi_{40}\rangle$ to $\vert\psi_{04}\rangle$.

\textbf{Theorem:} Let $n$ be even so that both
$\vert\psi_{n0}\rangle$ and $\vert\psi_{0n}\rangle$ are separable.
Entangled states $\vert\psi_{pq}\rangle$ ($pq\neq0$) are stable
under ILOs in the sense that either no ILO generated curve (given by
eq. (\ref{curve})) will pass through them or even if there is a
curve containing $\vert\psi_{pq}\rangle$ (for $t=1$ or $t=-1$),
probability of success will be decreasing to zero as
$\vert\psi(t)\rangle$ tends to $\vert\psi_{pq}\rangle$.

\textbf{Proof:} Let $c_{pq}$ and $c_{p'q'}$ ($n>p>p'>0$) be the only
non-vanishing coefficients so that the curve becomes the geodesic
connecting $\vert\psi_{pq}\rangle$ and $\vert\psi_{p'q'}\rangle$.
Since
$\vert\psi(t)\rangle\propto(1+t)^{p'}(1-t)^q[c_{pq}(1+t)^{p-p'}
\vert\psi_{pq}\rangle+c_{p'q'}(1-t)^{q'-q}\vert\psi_{p'q'}\rangle]$,
$t\rightarrow-1$ and $t\rightarrow1$ limits correspond to
$\vert\psi_{p'q'}\rangle$ and $\vert\psi_{pq}\rangle$ states,
respectively. The vectors vanish in these two limiting cases and
therefore probabilities for these events to occur go to zero.  If
more than two expansion coefficients are non-vanishing, it is not
possible to reach all eigenstates $\vert\psi_{pq}\rangle$ in the
sense that $\vert\psi_{pq}\rangle$ has a small enough neighborhood
which does not contain any points of the curve. Only those curves
where $p$ is the largest or smallest of the subscripts of
non-vanishing $c_{pq}$ coefficients pass through
$\vert\psi_{pq}\rangle$. This ends the proof.

The odd qubit $n=2m+1$ problem is equivalent to the even $n=2m$ case
in the sense that both are $d=m+1$ dimensional problems and there is
a one-to-one correspondence between the components of the vectors in
two spaces given by
\begin{equation}
c_{2(m-k),2k}\longleftrightarrow c_{2(m-k)+1,2k}(1+t),
\end{equation}
where $k=0,1,2,...,d$. The factor $1+t$ does not change the vector
since they are all to be normalized but probabilities do change.
While $\vert\psi_{0,2m}\rangle$ is a reachable and separable state,
the corresponding odd space partner $\vert\psi_{1,2m}\rangle$ is
entangled and cannot be approached arbitrarily closely since the
probability decays to zero for such processes.

Stability or unreachability properties are specific to entangled
eigenstates of $M^{\otimes n}$. For separable states probabilities
of being approached are in general nonzero. This is understandable
since local measurements are enough to collapse the whole wave
function into a separable one. For entangled states in different
equivalence classes conversion probability is clearly zero. In this
work, it is shown that, at least for FES entangled states, even
though the final state is in the same equivalence class as the
initial state, probability of success decays to zero as the final
state becomes nearer to the boundaries of the equivalence class.

In conclusion, a systematic method to classify FES $n-$qubit
entangled states has been presented. It has been shown that ILOs
result in a set of curves in the Hilbert space. Some entangled
states, namely eigenstates of FES ILOs, embedded in other entangled
states, have been found to be either totally unreachable, i.e., no
curves pass through them, or even if they are on a curve, the
probability decays to zero as they are approached. This observation
is important because it justifies SLOCC classification. As one
approaches to a boundary between two different classes,
probabilities get smaller and smaller. Since probability is a
continuous function, the same results must hold for states not
necessarily FES but in the vicinity of FES ones. Finally, FES
entangled $n-$qubit states given above are also good tests for
algebraic invariants proposed to distinguish SLOCC equivalence
classes \cite{19,20}. Even though general SLOCC classification is a
difficult problem, FES subspace classification is trivial and hence
one can start from this easy case to propose new invariants.

\section{Acknowledgements}
This work has been partially supported by the Scientific and
Technological Research Council of Turkey (T\"{U}B\.ITAK) under Grant
107T530. The author would like to thank \"{O}. Er\c{c}etin, G.
Karpat and C. Sa\c{c}l{\i}o\u{g}lu for helpful discussions and the
Institute of Theoretical and Applied Physics at Turun\c{c} where
part of this research has been done.

\end{document}